\documentclass[preprint,1p,11pt]{ISAS_IR}

\usepackage[english]{babel}
\usepackage{algorithmic}
\usepackage{amsthm}
\usepackage{ISASMatheMacros}
\usepackage{ISASTextMacros}
\usepackage{graphicx}
\usepackage{subfig}
\usepackage{url}

\begin{document}

\begin{frontmatter}

\title{The Kernel-SME Filter for Multiple Target Tracking}

\author{%
\textbf{Marcus Baum} and \textbf{Uwe D.~Hanebeck}\\
Intelligent Sensor-Actuator-Systems Laboratory (ISAS)\\
Institute for Anthropomatics\\
Karlsruhe Institute of Technology (KIT), Germany\\
{\tt marcus.baum@kit.edu, \tt uwe.hanebeck@ieee.org}
}

\begin{abstract}
We present a novel method  called Kernel-SME filter for tracking multiple targets  when the association  of the measurements to the  targets is unknown. The method is a  further development of  the Symmetric Measurement Equation (SME) filter, which removes the data association uncertainty of the original measurement equation with the help of a symmetric transformation. The underlying idea of the Kernel-SME filter is to construct a symmetric transformation by means of mapping the measurements to a Gaussian mixture. This transformation is scalable to a large number of targets and allows for deriving a   Gaussian state estimator  that has  a cubic time complexity in the number of targets.
\end{abstract}

\end{frontmatter}

\section{Introduction}
A main  challenge in multiple target tracking   \cite{Bar-Shalom2011} is that the association of measurements to targets is unknown. In this context, a variety of different multiple target tracking methods  has  been developed. For example, the \emph{Joint Probabilistic Data Association Filter (JPDAF)} \cite{Bar-Shalom2009} enumerates all feasible association hypotheses in order to compute a Gaussian approximation of the posterior density of the target states. Unfortunately, the number of possible association hypotheses grows exponentially with the number of targets so that  the tracking  of a large number of closely-spaced targets becomes a serious  challenge. The \emph{Probability Hypothesis Density (PHD) filter} \cite{Mahler2007,Vo2006}  maintains the   first moment of the multi-target posterior random set called PHD. By this means, association hypotheses are not explicitly enumerated, i.e., data association is performed implicitly. The PHD, however, contains significantly less information than the full joint state vector of all single targets, e.g., the correlations between targets are lost.

This article is about an  implicit data association approach  named \emph{Symmetric Measurement Equation (SME)} filter \cite{Kamen1992,Kamen1993}. The SME filter removes the data association uncertainty from  the original measurement equation using  a symmetric transformation. By this means, the combinatorial complexity of the data association problem can be bypassed. Unfortunately, existing SMEs suffer from strong nonlinearities  and lack an intuitive semantic so that  existing SME filters are not competitive to established approaches such as JPDAF and PHD filters.

In this article, we introduce the so-called Kernel-SME filter that can be seen as an extension of the SME approach. The basic idea is to define  a symmetric transformation that maps the set of  measurements to a function, i.e., a Gaussian mixture,  and deterministic sampling  of this function gives the symmetric transformation. In this manner, a data-dependent, i.e., nonparametric,  symmetric transformation is obtained. The Kernel-SME has an intuitive semantic and it is suitable for a large number of closely-spaced targets due to a cubic time complexity. Hence, the advantages of an implicit data association method are exploited  while having the full joint density of the multiple target state available. Additionally, there is an intriguing  connection to the PHD filter that renders the Kernel-SME filter to an in-between of the PHD filter and the JPDAF. Simulations demonstrate that the Kernel-SME filter may outperform  the PHD filter for a large number of targets.

\section{Problem Formulation}\label{sec:probform}
We consider the   tracking of multiple targets based on noisy  measurements, where the target-to-measurement association is unknown. 
Specifically, we make the following assumptions:
\begin{itemize}
 \item[A1] The number of targets is known and fixed.
 \item[A2] Each target gives rise to exactly one single  measurement per time instant.
 \item[A3] There are no false  measurements, i.e., each measurement originates from a target.
 \end{itemize}

The $n$-dimensional single target state vectors are denoted with $\targetr{k}[1], \ldots , \targetr{k}[N]$, where $k$ denotes the discrete time and $N$ is the number of targets. The joint target state    $\targetr{k} = \tvect{(\targetr{k}[1])^T, \ldots , (\targetr{k}[N])^T} \in \IR^{n\cdot N}$ comprises all single target states.

\subsection{Measurement Model}
At each time step $k$,   a    set of $N$ measurements  $ \{ \measr{k}[1],\ldots , \measr{k}[N] \}$   is available.   
Each measurement  is related to a single target through the linear measurement model
\begin{equation}\label{eqn:singletarget_meas}
\measr{k}[\pi_k(l)] =  \measma{k}[l]\targetr{k}[l]+\mnoise{k}[l] \enspace ,
\end{equation}
where $\pi_k \in \Pi_n$  is a permutation in the symmetric group  $\Pi_n$ that specifies the \emph{unknown} target-to-measurement assignment and $\mnoise{k}[l]$ is additive zero-mean white noise with  covariance matrix $\cov{v}[k][l]$.
The single target  measurement equations \Eq{eqn:singletarget_meas} can be composed to an overall measurement equation 
\begin{equation}\label{eqn:meas}
\underbrace{ \vect{\measr{k}[\pi_k(1)] \\ \vdots\\ \measr{k}[\pi_k(N)]}}_{=P_{\pi_k}(\measr{k})} = \underbrace{  \vect{ \measma{k}[1] &   &  \\  &  \ddots &   \\   &  &\measma{k}[N]}}_{= \measma{k}} \cdot  \underbrace{\vect{ \targetr{k}[1] \\  \vdots\\  \targetr{k}[N]} }_{=\targetr{k} }          + \underbrace{\vect{ \mnoise{k}[1] \\  \vdots\\  \mnoise{k}[N]} }_{=\mnoise{k} } \enspace ,
\end{equation}
where $      \measr{k}:=           \tvect{ (\measr{k}[1])^T,\ldots , (\measr{k}[N])^T }$   and $P_{\pi_k}(\measr{k})$ permutes the single measurements in  $\measr{k}$ according to $\pi_k$.

\subsection{System Model}
 The temporal evolution of a single target is specified by a linear motion model 
 \begin{equation}\label{eqn:singletarget_sys}
 \targetr{k+1}[l] = \sysma{k}[l] \targetr{k}[l]  + \sysnoise{k}[l]   \enspace , 
 \end{equation}
where $\sysma{k}[l] $ is the system matrix and $\sysnoise{k}[l]$ is additive white noise with covariance matrix $\cov{w}[k][l]$.
 The single target motion models   \Eq{eqn:singletarget_sys}  can be composed as 
\begin{equation}
\underbrace{\vect{ \targetr{k+1}[1] \\  \vdots\\  \targetr{k+1}[N]} }_{=\targetr{k+1} } = \underbrace{ \vect{ \sysma{k}[1] &   &  \\  &  \ddots &   \\   &  &\sysma{k}[N]}}_{:= \sysma{k}} \cdot  \underbrace{\vect{ \targetr{k}[1] \\  \vdots\\  \targetr{k}[N]} }_{=\targetr{k} }          + \underbrace{\vect{ \sysnoise{k}[1] \\  \vdots\\  \sysnoise{k}[N]} }_{=\sysnoise{k} } \enspace .
\end{equation}

\subsection{Recursive Gaussian Estimator}
We aim at   a recursive Gaussian state estimator for the  multi-target state vector $\targetr{k}$, i.e., a Gaussian approximation of the  posterior probability density function for $\targetr{k}$ given the measurements $\mathbf{Y}_k := \{ \measr{1}, \ldots , \measr{k} \} $ 
 
\begin{equation}\label{eqn:state_estimate_gaussian}
  p(\targetd{k}\, |\,  \mathbf{Y}_k  )  = \Gauss{\targetd{k}}{\mean{x}[k]}{\cov{x}[k]}  \enspace
\end{equation}
is to be computed, where  $\mean{x}[k]$  is the mean  and   $\cov{x}[k]$ the covariance matrix of the  Gaussian.

The time update step determines  $p(\targetd{k}\, |\,  \mathbf{Y}_{k-1}  )=\Gauss{\targetd{k}}{\mean{x}[k|k-1]}{\cov{x}[k|k-1]} $ based on the previous density $p(\targetd{k-1}\, |\,  \mathbf{Y}_{k-1}  )$. Due to the linear system model, the prediction can be performed according the Kalman filter formulas
\begin{eqnarray} 
 \mean{x}[k|k-1] &=& \sysma{k}   \cdot \mean{x}[k-1] \enspace,  \text{ and}\\
 \cov{x}[k|k-1] & =&     \sysma{k} \cov{x}[k-1] (\sysma{k})^T+ \cov{w}[k] \enspace .
\end{eqnarray}
 In the measurement update step, the prediction $\Gauss{\targetd{k}}{\mean{x}[k|k-1]}{\cov{x}[k|k-1]} $  is updated with  the stacked measurement vector $\measr{k}$. How to perform the measurement update under incorporation of the data association uncertainty    is the objective this  article.

\section{SME-Filter}
This section is about  the \emph{Symmetric Measurement Equation (SME)} filter as introduced by Kamen \cite{Kamen1992,Kamen1993}. The basic idea of the SME filter is to remove the  association uncertainty $\pi_k$ from  the measurement equation \Eq{eqn:meas} by applying a symmetric transformation to the measurement vector.
\begin{Definition}
A transformation $S(\measr{k} )$ of the measurement vector  $\measr{k}$ with  $S\hspace{-0.06cm}: \IR^{N \cdot  n }\rightarrow \IR^{N_a} $ is called symmetric if
 \begin{equation}
  S(\measr{k} ) = S(P_{\pi}(\measr{k}))
 \end{equation}
for all $\pi \in \Pi_N$.
\end{Definition}
\begin{Remark}
 Of course, the symmetric transformation should not remove information, i.e., it should be injective up to permutation. 
\end{Remark}

\begin{Example} The \emph{Sum-Of-Powers} \cite{Kamen1992,Kamen1993,Leven2009,Leven2004} transformation for two targets and one-dimensional measurements $ {\rv{y}}_{k}^1$ and $ {\rv{y}}_{k}^2$  is given by 
 $$S\(\tvect{\rv{{y}}_{k}^1,\rv{{y}}_{k}^2 } \)= \tvect{ \rv{ {y}}_{k}^1 +\rv{{y}}_{k}^2 ,    \(\rv{{y}}_{k}^1\)^2 +\(\rv{{y}}_{k}^2\)^2        }\enspace . $$
 \end{Example}
 
The application of a  symmetric function  $S$ to  \Eq{eqn:meas} yields
\begin{equation}\label{eqn:transformed_measeqn}
  \pmeasr{k}:=\underbrace{S(P_{\pi_k}(\measr{k}) )}_{= S(\measr{k})}= S(   \measma{k} \cdot     \targetr{k}  + \mnoise{k}  ) \enspace ,
 \end{equation}
 where $\pmeasr{k}$  is a pseudo-measurement constructed from original measurement vector $\measr{k}$. The pseudo-measurement $\pmeasr{k}$ can be determined without knowing $\pi_k$ due to the symmetry property of $S$. Hence, the data association uncertainty has been removed, however, instead a nonlinear  measurement equation is obtained. Based on the nonlinear measurement equation \Eq{eqn:transformed_measeqn}, nonlinear Bayesian state estimators such as the Extended Kalman Filter (EKF) or Unscented Kalman Filter (UKF) \cite{Leven2009,Leven2004} can be used for performing inference.

Although the SME approach is a very neat way for dealing with data association uncertainties, it comes with some challenges:
\begin{enumerate}
\item[1.)]  The generalization of existing symmetric transformations, i.e., the   \emph{Sum-Of-Powers} and  \cite{Kamen1992,Kamen1993,Leven2009,Leven2004,Muder1993},  to   states  with dimension larger than 1 is nontrivial due to the so-called ghost target problem \cite{Leven2009,Leven2004} resulting from non-injective transformations. As a consequence,  tedious and highly nonlinear symmetric functions  that  have no intuitive, physical meaning are obtained. Additionally, these  symmetric transformations are  unsuitable for larger target numbers as the order of the involved polynomial increases with the number of targets, i.e., for $10$ targets polynomials up to order $10$ are required.
 \item[2.)]  Due to 1.), the resulting nonlinear estimation problem is very difficult. As there is non-additive Gaussian noise  in  \Eq{eqn:transformed_measeqn}, the  EKF cannot be applied directly and an approximate measurement equation with additive noise has to be derived first.  The derivation of the additive noise term is usually complicated and time-consuming. Besides,  Linear Regression Kalman Filters (LRKFs) such as the UKF \cite{Leven2009,Leven2004} do  not give satisfying results due to the strong  nonlinearities and numerical instabilities.
\end{enumerate}

\section{Kernel-SME Filter}
The basic idea of the Kernel-SME filter is to  interpret the  measurements   as the parameters of a    function, where the function is  a sum of kernel functions that are placed at the measurement locations. We focus on Gaussian kernels, nevertheless other types of kernels may also be reasonable.

\begin{figure} 	   
	\centering
	\includegraphics[ ]{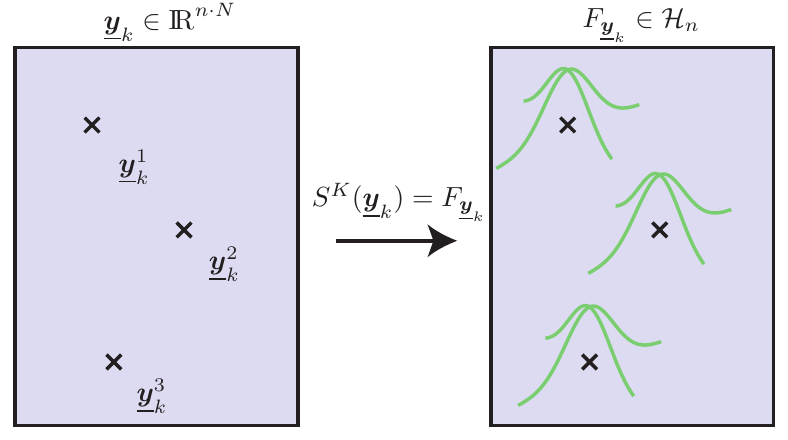}			
	\caption{Illustration of the Kernel-SME.\label{fig:example}}
	\end{figure}
\begin{Definition}[Kernel Transformation] Let $\mathcal{H}^N_n$ denote the space of all $n$-dimensional Gaussian mixtures with $N$ components. The kernel transformation  $S^K : \IR^{N\cdot n} \rightarrow \mathcal{H}_n^N$, which maps $\measr{k} \in \IR^{N\cdot n} $ to a function $F_{\measr{k}} \in\mathcal{H}_n^N$,   is defined as 
 \begin{equation}\label{eqn:kernelmapping}
  S^K(\measr{k} ) =             F_{\measr{k}} \enspace \text{with} 
 \end{equation}
 \begin{equation}\label{eqn:kernelmapping2}
             F_{\measr{k}}(\vec{z}) = \sum_{l=1}^{N} \Gauss{\vec{z}}{\measr{k}[l]}{\cov{K}} \enspace ,
 \end{equation} 
  where $\Gauss{\vec{z}}{\measr{k}[l]}{\cov{K}}$ is a Gaussian kernel located at $\measr{k}[l]$ with kernel width $\cov{K}$.
\end{Definition}
\begin{Remark}The transformation \Eq{eqn:kernelmapping} is symmetric due to 
$F_{\measr{k}}(\vec{z}) =  F_{P_{\pi}(\measr{k})}(\vec{z})$ for all $\vec{z} \in \IR^n $.
Furthermore, \Eq{eqn:kernelmapping} is injective up to permutation due to the identifiable of the parameters of a Gaussian mixture density  \cite{Yakowitz1968}. Hence, there is no ghost target problem.
\end{Remark}

The transformation  \Eq{eqn:kernelmapping}  has an intuitive semantic: The set of measurements is interpreted as a continuous image, i.e., high values of  $F_{\measr{k}}(\vec{z})$  indicate a high measurement concentration.  Of course, the choice of a suitable kernel width $\cov{K}$ in  \Eq{eqn:kernelmapping}  is essential.  It should be chosen similar to  the measurement noise covariance  in order to ensure that   the support of the kernel covers the    noise-free measurement.

The following theorem describes  an insightful, inherent relationship between  the transformed measurements $S^K(\measr{k} )$  and the   PHD of the stacked measurements $\measr{k}$ that is  defined as
   \begin{eqnarray}
 D_{\measr{k}}(\meas{}) &:=&  \sum_i p_{\meas{k}[i]}(\meas{}  ) \label{eqn:phd} \enspace ,
\end{eqnarray}
where $D_{\measr{k}}(\meas{})$  is the PHD function and   $p_{\measr{k}[i]}(\meas{}  )$ is the probability density of $\measr{k}[i]$.
 
 \begin{Theorem}The expected kernel transformation of the  stacked measurements $\measr{k}$ coincides with the convolution of the 
 PHD with the kernel, i.e., $\Exp{S^K(\measr{k} )} =  \int D_{\measr{k}}(\vec{s}) \cdot        \Gauss{\vec{s}}{\vec{z}}{\cov{K}} \;     \mathrm{d}\vec{s} $
 \end{Theorem}
 \begin{Proof} According to \cite{CDC12_Baum}, the following holds
  \begin{multline*}
 \Exp{ F_{\measr{k}}(\vec{s})}= \int     F_{\meas{k}}(\vec{s}) p(\meas{k} ) \, \mathrm{d}\meas{k}
 =    \int  \int  \sum_i \delta(  \vec{t}- \meas{k}[i]  ) \cdot  \Gauss{\vec{t}}{\vec{s}}{\cov{K}}  \, \mathrm{d} \vec{t}  \,         p(\meas{k} ) \, \mathrm{d}\meas{k} \\ 
  =    \int  \int  \sum_i \delta(  \vec{t}- \meas{k}[i]  )   \,         p(\meas{k} ) \, \mathrm{d}\meas{k}  \cdot \Gauss{\vec{t}}{\vec{s}}{\cov{K}}  \, \mathrm{d} \vec{t} 
  =    \int  D_{\measr{k}}(\vec{t}) \cdot \Gauss{\vec{t}}{\vec{s}}{\cov{K}}   \, \mathrm{d} \vec{t} \enspace .
 \end{multline*}
\end{Proof}

 In order to apply  standard nonlinear estimation techniques for determining the estimate  \Eq{eqn:state_estimate_gaussian}, we propose to evaluate the function $F_{\measr{k}}(\vec{z})$ at specific test vectors $ \testvec{k}[1], \ldots,      \testvec{k}[N_a] $, i.e., we define a  discretized version of \Eq{eqn:kernelmapping} as follows
  \begin{equation}\label{eqn:kernelmapping_discr}
  S^K_{\testvec{k}[1], \ldots,      \testvec{k}[N_a]}(\measr{k}  ) =    \vect{F_{\measr{k}}(\testvec{k}[1]) \\ \vdots \\  F_{\measr{k}}(\testvec{k}[N_a])  }
 \end{equation}
 How to choose the number and locations of the test vectors is discussed in  \Sec{sec:testvectors}.

 The application of \Eq{eqn:kernelmapping_discr} to \Eq{eqn:meas} gives the following symmetric measurement equation
  \begin{equation}\label{eqn:transformed_measeqn2}
  \pmeasr{k}= S^K_{\testvec{k}[1], \ldots,      \testvec{k}[N_a]}(\measr{k} ) = S^K_{\testvec{k}[1], \ldots,      \testvec{k}[N_a]}(   \measma{k} \cdot     \targetr{k}  + \mnoise{k}  ) \enspace, 
 \end{equation} 
 where  $\pmeasr{k}$  is the pseudo-measurement.

 We derive a   \emph{Linear Minimum Mean Squared Error (LMMSE)} estimator \cite{Bar-Shalom2002}. For a given prediction of the state  $\mean{x}[k|k-1]$  with  estimation error  $\cov{x}[k|k-1]$, the updated estimate $\mean{x}[k]$  and   $\cov{x}[k]$    according to \Eq{eqn:transformed_measeqn2} is given by the Kalman filter formulas
 \begin{eqnarray}
  \mean{x}[k]  &=&  \mean{x}[k|k-1]    + \cov{xs}[k] (\cov{ss}[k])^{-1} \left(\pmeasr{k}-\mean{s}[k]  \right) \enspace , \text{ and} \label{eqn:kalmanupdate_mean}   \\
     \cov{x}[k]       &=& \cov{x}[k|k-1]-  \cov{xs}[k](\cov{ss}[k])^{-1} \cov{sx}[k], \label{eqn:kalmanupdate_cov}
   \end{eqnarray}
 where 
 \begin{itemize} 
  \item  $\mean{s}[k] $ is the predicted pseudo-measurement,
  \item $\cov{xs}[k]$ is the covariance  between the state vector $\targetr{k}$ and the pseudo-measurement $\pmeas{k}$ , and 
  \item $\cov{ss}[k]$ is the variance of the pseudo-measurement $\pmeas{k}$.
 \end{itemize}
  
 Intuitively, the above filter minimizes the kernel distance between the  PHD  of the predicted measurements  and the measurements. In this context, see also \cite{CDC12_Baum}.
   
 Closed-form expressions for the above moments are derived in \Sec{sec:closedform}. With these expressions, the computational complexity  of the Kalman filtering update \Eq{eqn:kalmanupdate_mean} and  \Eq{eqn:kalmanupdate_cov} is only  cubic in the number of targets: The mean $\mean{s}[k]$  can be computed in quadratic time,  and  both the cross-covariance matrix $\cov{xs}[k]$ and covariance matrix $\cov{ss}[k]$ have a cubic time complexity (see \Sec{sec:closedform}). Hence, the overall  time complexity of the measurement update is cubic.

\subsection{Selecting the Test Vectors}\label{sec:testvectors}
The locations of the test vectors are crucial. Fortunately, there is an intuitive interpretation: As the test vectors  can be seen as deterministic samples of the Gaussian mixture     \Eq{eqn:kernelmapping2}, there is a strong relationship to deterministic sampling problems  that occur for example in the UKF \cite{Julier_UnscentedFiltering}. Due to this analogy, we propose to add  $2 \cdot n$ test vectors for each   Gaussian component   $\Gauss{\vec{z}}{\measr{k}[l]}{\cov{K}}$  in  \Eq{eqn:kernelmapping2}   according to the UKF \cite{Julier_UnscentedFiltering}, i.e.,  $N_a = 2 \cdot n\cdot N$ with 
  \begin{eqnarray}
    \testvec{k}[l+i-1]           &=& \measr{k}[l]  + \(\sqrt{{n \cov{K}}}\)_i  \enspace ,  \text{ and } \\
    \testvec{k}[l+2(i-1)]           &=& \measr{k}[l]  - \(\sqrt{{n \cov{K}}}\)_i  \enspace    
  \end{eqnarray}
for $i= 1,\ldots,N$ and $l= 1,\ldots,n$,  where $\big(\sqrt{{n \cov{K}}}\big)_i$ denotes the $i$-th column of $\sqrt{{n\cov{K}}}$. For  diagonal $\cov{K}$,  the test points lie on the principal components   of $\cov{K}$.

\subsection{Closed-Form Expressions}\label{sec:closedform}
The moments required for the measurement update step can be calculated in closed form. Essentially, the derivations are straightforward as they can be performed with the help of the Kalman filtering formulas. For this purpose, we define the abbreviation
\begin{eqnarray*}
 P_{i,l} &:=&  \Gauss{\testvec{k}[i]}{\measma{k}[l]\mean{x}[l]}{ \measma{k}[l]    \cov{x}[k|k-1]    (\measma{k}[l])^T+ \cov{v}[k] + \cov{K} } \enspace .
\end{eqnarray*}
The mean of the predicted pseudo-measurement  $\mean{s}[k] =  \tvect{\mean{s_1}[k], \ldots ,\mean{s_{N_a}}[k] }  $ is
\begin{multline}
\mean{s}[k][i]  =   \Exp{F_{\measr{k}}(\testvec{k}[i])\, | \, \mathcal{Y}_{k-1}} 
=     \sum_{l=1}^{N} \int  \Gauss{\testvec{k}[i]}{\measma{k}[l]\targetd{k}[l]+\mnoised{k}[l]}{\cov{K}} \cdot \\ 
\Gauss{\targetd{k}}{\mean{x}[k|k-1]}{\cov{x}[k|k-1]}  \cdot \Gauss{\mnoised{k}}{\vec{0}}{\cov{v}[k]}  \, \mathrm{d}  \targetd{k}[l] \, \mathrm{d}  \mnoised{k}[l]          
=\sum_{l=1}^{N} P_{i,l} \enspace .
\end{multline}
The cross-covariance matrix between the multi-target state vector and the pseudo-measurement  becomes
$\cov{xs}[k]  =  \vect{\cov{xs_1}[k], \ldots ,\cov{xs_{N_a}}[k] }$   with 
\begin{multline}
 \cov{xs_i}[k] =   \underbrace{\Exp{ \targetr{k} \cdot  F_{\measr{k}}(\testvec{k}[i])\, | \, \mathcal{Y}_{k-1}}}_{(*)}  - \mean{x}[k] \cdot  \mean{s}[k][i] \enspace ,  \text{ where }\\ 
(*)  =     \sum_{l=1}^{N} \int \targetd{k} \cdot \Gauss{\testvec{k}[i]}{\measma{k}[l]\targetd{k}[l]+\mnoised{k}[l]}{\cov{K}} \cdot 
 \Gauss{\targetd{k}}{\mean{x}[k|k-1]}{\cov{x}[k|k-1]}  \cdot \Gauss{\mnoised{k}}{\vec{0}}{\cov{v}[k]}  \, \mathrm{d}  \targetd{k} \, \mathrm{d}  \mnoised{k}[l]           \\
= N \mean{x}[k|k-1] +\sum_{l=1}^{N}   P_{i,l}  \mat{K}_{k}^l (\testvec{k}[i]-  \measma{k}[l]\mean{x_l}[k]) \text{ and } \\
 \mat{K}_{k}^l  =   \tvect{ \cov{x_1 x_l}[k|k-1] ,  \ldots,   \cov{x_Nx_l}[k|k-1] }\measma{k}[l] \cdot 
 \(\measma{k}[l] \cov{x}[k|k-1][l](\measma{k}[l])^T +   \cov{K}[k] +\cov{v}[k]\)^{-1} \enspace .
 \end{multline}
The   covariance matrix of the predicted  pseudo-measurement 
$\cov{ss}[k]  =   (\cov{s_is_j}[k])_{i,j = 1, \ldots,N_a}$  can be calculated with 
\begin{multline}
\cov{s_is_j}[k] =  \underbrace{ \Exp{   F_{\measr{k}}(\testvec{k}[i]) \cdot F_{\measr{k}}(\testvec{k}[j])        \, | \, \mathcal{Y}_{k-1}}}_{(**)}  -    \mean{s}[k][i] \cdot  \mean{s}[k][j] \enspace , \text{ where }\\ 
(**) =     \sum_{l=1}^{N}  \sum_{m=1}^{N} \int  \Gauss{\testvec{k}[i]}{\measma{k}[l]\targetd{k}[l]+\mnoised{k}[l]}{\cov{K}} \cdot  
\Gauss{\testvec{k}[j]}{\measma{k}[m]\targetd{k}[m]+\mnoised{k}[m]}{\cov{K}}   \cdot
 \Gauss{\targetd{k}}{\mean{x}[k|k-1]}{\cov{x}[k|k-1]}  \cdot \\ \Gauss{\mnoised{k}}{\vec{0}}{\cov{v}[k]}  \, \mathrm{d}  \targetd{k} \, \mathrm{d}  \mnoised{k}[l]        
= \Bigg( \sum_{l=1}^{N}   P_{i,l}        \sum_{m=1, m\neq l}^{N}     P_{j,m}     \Bigg) 
+  \sum_{l=1}^{N}     \Gauss{\testvec{k}[i]}{\testvec{k}[j] }{0.5\cov{K}}   \cdot \\  
\Gauss{\tfrac{1}{2}(\testvec{k}[i]+\testvec{k}[j] )}{\measma{k}[l]\mean{x_l}[k]}{ \measma{k}[l]    \cov{x}[k|k-1]    (\measma{k}[l])^T+ \cov{v}[k]} \enspace . 
\end{multline}

\section{Evaluation}
\begin{figure}
\centering
\subfloat[Initial target positions (big dots) and an example random walk taken by the targets (small dots).]{
 \includegraphics[width=4cm]{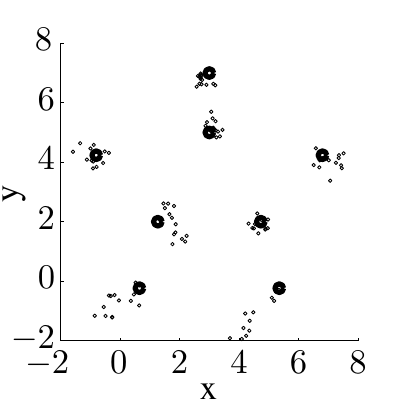}
 \label{fig:sim_initial}
}  
\hspace{0.5cm}
\subfloat[Example measurements.]{
 \includegraphics[width=4cm]{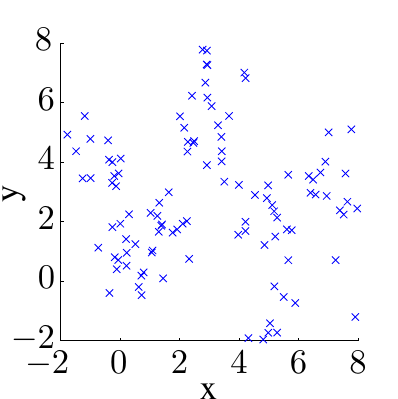}
 \label{fig:sim_meas}
}  
\hspace{0.5cm}
\subfloat[Mean OSPA Error: Kernel-SME (blue) and GM-PHD (magenta) filter.]{
\includegraphics[width=4cm]{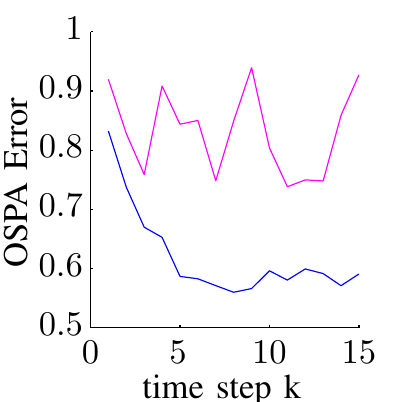}
 \label{fig:sim_mospa}
}  
\caption{Simulations: Setting and results for the  first $15$ time steps.}
\end{figure}

The performance of the Kernel-SME filter is demonstrated with respect to the Gaussian mixture implementation of the  PHD  filter  (GM-PHD)  \cite{Vo2006}. For this purpose, eight two-dimensional targets that evolve  according to a random walk model are considered (see \Fig{fig:sim_initial}), i.e., $N=2$ and  $n=8$ with parameters $\measma{k}[i] = \sysma{k}[i]= \mat{I}_{2} $, $\cov{v}=0.1$, and  $\cov{w}=0.1$, where $\mat{I}_{2}$ is the identity matrix of dimension $2$. The measurement noise is rather high compared to the distance of the targets. The first estimate is initialized with the covariance matrix $\cov{x}[0]= 0.5 \cdot \mat{I}_{16} $ and the mean $\mean{x}[0]$ is sampled randomly from $\Gauss{\tilde{\vec{x}}_0}{\vec{0}}{\cov{x}[0]  }$, where $\tilde{\vec{x}}_0$ denotes the true target position at time instant $0$. The GM-PHD filter maintains a Gaussian mixture with $50$ components   in order to represent the PHD. The parameters for the Gaussian mixture reduction have been optimized for the best results and the mixture components with the largest weights serve as point estimates for the single targets. As the PHD filter itself does not maintain target labels, the performance of both filters is assessed with the  \emph{Optimal Sub-Pattern Assignment (OSPA)} metric \cite{Schuhmacher2008} that ignores target labels. The averaged  OSPA distance over $30$ Monte Carlo runs is depicted \Fig{fig:sim_meas}. The Kernel-SME filter significantly outperforms the PHD filter in this scenario. The reason is that the PHD tends to merge closely-spaced targets. The simulations demonstrate that the Kernel-SME filter is  advantageous in particular settings. However, please note that the PHD filter is more general than the presented Kernel-SME filter version, e.g., the PHD filter is capable of estimating the number of targets.

\section{Conclusions}
This article presented a novel type of SME filter that is based on a mapping from  the set of measurements to a Gaussian mixture. Intuitively, the filter recursively minimizes the kernel distance between the measurements and the PHD of the predicted measurements. By this means, shortcomings of existing SME approaches are remedied so that   Kernel-SME filter   is a serious alternative to traditional tracking algorithms such as JPDAF and PHD filters. The  Kernel-SME filter  is in particular advantageous for a large number of closely-spaced targets. Future work focuses on  extending the Kernel-SME to clutter measurements and detection probabilities  (see A2 and A3 in \Sec{sec:probform}).

\bibliographystyle{IEEEtran}
\bibliography{Literatur,ISASPublikationen,ISASPublikationen_laufend}

\end{document}